\documentclass[conference]{IEEEtran}
\IEEEoverridecommandlockouts
\usepackage[bookmarks=false]{hyperref}
\usepackage[T1]{fontenc}
\usepackage{lipsum}
\usepackage{subfig}
\usepackage{hyperref}
\usepackage{url}
\usepackage{amssymb}
\usepackage{rotating}
\usepackage{multirow}
\usepackage{xspace}
\usepackage{array}
\usepackage{amssymb}
\usepackage{amsfonts}
\usepackage{graphicx}
\usepackage{epstopdf}
\usepackage{amsmath}
\usepackage{booktabs}
\usepackage{epigraph}
\usepackage{listings}
\usepackage{color}
\usepackage{wasysym}
\usepackage{float}
\usepackage{amsmath}
\usepackage{pifont}
\usepackage[table,xcdraw]{xcolor}

\usepackage{booktabs}
\usepackage{listings}
\usepackage{color}
\usepackage{fontawesome}
\usepackage{threeparttable}

\def\BibTeX{{\rm B\kern-.05em{\sc i\kern-.025em b}\kern-.08em
    T\kern-.1667em\lower.7ex\hbox{E}\kern-.125emX}}
    
\begin{document}

\title{How Successful Are Open Source Contributions From Countries with Different Levels of Human Development?}

\author{\IEEEauthorblockN{Leonardo B. Furtado}
\IEEEauthorblockA{Federal University of Par\'a\\
Bel\'em, Par\'a, Brazil\\
srleonardofurtado@gmail.com}
\and
\IEEEauthorblockN{Bruno Cartaxo}
\IEEEauthorblockA{Federal Institute of Pernambuco\\
Paulista, Pernambuco, Brazil\\
email@brunocartaxo.com}
\and
\IEEEauthorblockN{Christoph Treude}
\IEEEauthorblockA{University of Adelaide\\
Adelaide, Australia\\
christoph.treude@adelaide.edu.au}
\and
\IEEEauthorblockN{Gustavo Pinto}
\IEEEauthorblockA{Federal University of Par\'a\\
Bel\'em, Par\'a, Brazil\\
gpinto@ufpa.br}
}

\maketitle

\begin{abstract}
Are Brazilian developers less likely to have a contribution accepted than their peers from, say, the United Kingdom? In this paper we studied whether the developers' location relates to the outcome of a pull request. We curated the locations of 14k contributors who performed 44k pull requests to 20 open source projects. Our results indeed suggest that developers from countries with low human development indexes (HDI) not only perform a small fraction of the overall pull requests, but they also are the ones that face rejection the most.
\end{abstract}


\section{Introduction}

%The development process is inherently a distributed activity. Given the presence of social coding platforms, developers world-wide can contribute to any open source project, if technical abilities and time constraints were not an issue.

Developers in open source software (OSS) projects must make decisions on contributions made by other community members, such as whether or not to accept a pull request. Previous studies have shown that factors such as gender and community status may influence the chances of contributions being accepted~\cite{Tsay:2014:IST:2568225.2568315}.

In this paper we studied whether developers based in countries with low human development are less likely to succeed in contributing to OSS projects. We used the Human Development Index (HDI) to measure the human development of a country.  HDI measures three dimensions of human development: health, education, and income per capita. According to the United Nations Development Programme\footnote{http://hdr.undp.org/en/content/human-development-index-hdi}, ``the health dimension is assessed by life expectancy at birth, the education dimension is measured by mean of years of schooling for adults aged 25 years and more and expected years of schooling for children of school entering age. The standard of living dimension is measured by Gross Domestic Product (GDP) per capita.'' HDI is also widely used by the United Nations (UN)~\cite{un_hdi_report_2019} and many other international organizations.  

%We, however, are not interested in studying biases towards less developed countries; our intention is to draw a quantitative landscape of this phenomenon.

To conduct this work, we analyzed 44,630 pull requests performed by 14,133 contributors into 20 well-known and well-studied OSS projects. %We created a tool to extract the contributors' locations. 
Our investigation suggests that, indeed, developers based in low HDI locations perform fewer pull requests and, proportionally, are the ones with the highest rejection rates.

\section{Method}

To conduct our work, we chose OSS projects that are (1) long-lived (i.e., more than two years of historical records), (2) popular (i.e., more than 5,000 stars on GitHub), (3) well-studied (i.e., studied in other research works), (4) diverse (i.e., in terms of their domain), and (5) active (i.e., more than 1,000 pull requests submitted). We then manually selected a few OSS projects that met these criteria. They are: \textsc{atom/atom}, \textsc{d3/d3}, \textsc{php/php-src}, \textsc{Microsoft/vscode}, \textsc{django/django}, \textsc{mongodb/mongo}, \textsc{ionic-team/ionic}, \textsc{python/cpython}, \textsc{facebook/react}, \textsc{mozilla-mobile/firefox-ios}, \textsc{apple/swift}, \textsc{Homebrew/brew}, \textsc{scikit-learn/scikit-learn}, \textsc{laravel/laravel}, \textsc{angular/angular}, \textsc{zulip/zulip}, \textsc{facebook/react-native}, \textsc{spyder-ide/spyder}, \textsc{tensorflow/tensorflow}, and \textsc{vuejs/vue}.

For each OSS project, we crawled contributors' (e.g., names, GitHub handles, location, etc.) and contributions' (e.g., pull requests performed, pull request status, etc.) data. Overall, we obtained data from 16,836 contributors and 96,592 contributions. We applied some criteria for analyzing pull requests data:

\begin{itemize}
    \item First, we excluded pull requests that were integrated by the submitters themselves, thus excluding 22,356 pull requests. 
    \item Second, we identified contributors with organizational email addresses and we excluded their pull requests. We did this because these developers can work for companies that support these projects and have a large stake in sending pull requests, most of which are more likely to be accepted. This excluded other 5,544 pull requests. 
    \item Third, we excluded pull requests from contributors who are part of the project organization or who are part of some organization that funds this project. To find the names of these organizations we inspected project pages and looked for backers or funders pages. This led to the exclusion of 18,823 pull requests. 
\end{itemize}

After these procedures, we were left with 49,869 pull requests from 15,654 contributors.

Since location is not a mandatory field on GitHub, we observed that not all contributors have filled it. We discarded contributors that did not provide their location. Moreover, on GitHub, the location field is a free text form; therefore, GitHub users can fill it with any information. We created a tool that matches the textual information provided in the location field with a location database, curated by \texttt{\url{simplemaps.com}} (Simplemaps for simplicity). Simplemaps is a database that provides the name of cities, states, countries, and other geographical information. According to its website, they ``built it from the ground up using authoritative sources such as the NGIA, US Geological Survey, US Census Bureau, and NASA.''\footnote{https://simplemaps.com/data/world-cities} 
%Moreover, they also provide a free version of their database, which we used in this study.
Although Simplemaps provides a comprehensive database, some adjustments were still needed. For instance, since we perceived that some GitHub users fill their locations with well-known acronyms (e.g., developers often use NY and NYC to mean New York City), we had to enrich the database with them. 
Using this approach, we were able to categorize 14,133 (90\%) of contributors that filled the location field. We discarded the 1,521 GitHub contributors for which we could not infer their location. 

Regarding the contributions, we focused only on \emph{closed} pull request, due to our interest in analyzing the relationship between the contributors' location and the acceptance/rejection of the pull request. Therefore, we had to rely on pull requests that have already passed through the code review process. A total of 44,630 pull requests were then selected for analysis. These pull requests were submitted between September 2010 and September 2019 (when we collected data). 

Regarding the countries' population and HDI, we used the UN database\footnote{http://hdr.undp.org/en/data}. We adopted the same four level HDI stratification (very high, high, medium, and low) that UN traditionally uses in their reports \cite{un_hdi_report_2019}. We considered the year of 2018, the most recent data available. 

Our data and tools are available at: \url{https://github.com/LeonardoFurtado/github-user-informations-collector}.

\section{Results}

\noindent
\textbf{Contributions based on the location.} %Figure~\ref{fig:overall} shows the world-wide distribution of the pull requests performed in the studied projects. As we can see,
Overall, developers from the United States, United Kingdom, and Germany performed the highest number of contributions (20,731 out of the 44,630 analyzed PRs), regardless if the contribution was accepted or not. In particular, contributors based in the United States are by far the most active ones in this regard, performing 14,795 PRs (33\% of the total contributions). Table~\ref{tab:prs} summarizes the top-20 locations of the developers that contributed the most to our studied projects. %One interesting observation from this table is that only three of the 20 developers' locations are below the equator line (they are: New Zealand, Australia and Brazil), suggesting that there is ample opportunity for developers from these locations to onboard in OSS projects. 
If we consider the countries' HDI, only four (20\%) out of the top-20 are not at the Very High HDI level (0.800--1.000), namely China, India, Brazil, and Ukraine. This lack of representativeness for lower HDI countries becomes even more evident when we consider the top-20 countries, but this time by the number of PRs per country population. All of them have Very High HDI levels. This shows that, although countries like China, India, and Brazil are in the top-20 when considering the absolute number of PRs, this is probably due to their large populations. In terms of individual work, Canada is the location that has the highest ratio of PRs per contributor (4.15 PRs/contributor), followed by France (3.77 PRs/contributor), and United States (3.72 PRs/contributor). On the other hand, Latin America-based and Africa-based developers are significantly less active than their peers from North America, Europe, and Oceania. Latin American developers performed only 1,183 (2\%) of the total contributions in our dataset. %More precisely, only developers based on 15 African countries (among the total 55 countries) performed contributions (265 pull requests).

\begin{table*}[]
  \caption{Number of pull requests performed per developers' location.}
  \centering
  \begin{threeparttable}
  \begin{tabular}{lrrrrr|r}
    \hline
    Country         & \# PRs & \# PRs/Pop.M.    & \# Contr. & \# PRs/Contr. & Acc. & HDI \\ \hline
    USA             & 14,795 & 12.91            & 4,223     &3.50           &  44.45\%   & 0.920    \\
    UK              & 3,179  & 13.22            & 887       &3.58           &  48.13\%   & 0.920    \\
    Germany         & 2,757  & 11.01            &915        & 3.01          &  44.98\%   & 0.939    \\
    India           & 2,590  & 0.51             &696        & 3.72          & 35.71\%    & 0.647    \\
    Canada          & 2,301  & 14.93            &554        & 4.15          & 35.77\%    & 0.922     \\
    France          & 2,053  & 8.38             &545        & 3.77          & 45.93\%    & 0.891     \\
    China           & 1,860  & 0.53             &758        & 2.45          & 39.09\%    & 0.758     \\
    Australia       & 1,212  & 14.62	        &364        & 3.33          &  46.70\%   & 0.938     \\
    Japan           & 1,171  & 2.96 	        &377        & 3.11          &  51.24\%   & 0.915     \\
    Netherlands     & 942    & 21.64            &370        & 2.55          &  32.91\%   & 0.933     \\
    Russia          & 846    & 2.23             &325        & 2.60          &  39.01\%   & 0.824     \\
    Brazil          & 780    & 1.62             &339        & 2.30          &  32.56\%   & 0.761     \\
    Poland          & 650    & 5.73             &217        & 3.00          &  25.69\%   & 0.872     \\
    New Zealand     & 576    & 33.40            &157        & 3.67          &  39.93\%   & 0.921     \\
    Sweden          & 544    & 19.70            &197        & 2.76          &  40.07\%   & 0.937     \\
    Switzerland     & 432    & 19.06            &162        & 2.67          &  45.83\%   & 0.946    \\
    Taiwan          & 362    & 4.77             &113        & 3.20          &  44.20\%   & 0,911     \\
    Spain           & 352    & 3.38             &158        & 2.23          &  27.27\%   & 0.893     \\
    Ukraine         & 338    & 3.44             &152        & 2.22          &  33.43\%   & 0.750     \\
    South Korea     & 325    & 2.40             &123        & 2.64          &  59.69\%   & 0.906      \\ \hline
  \end{tabular}
      \begin{tablenotes}
        \item PRs: Pull requests; Pop.M: Country Population (in Millions); Contr: Contributors; Accept: Acceptance Ratio.
    \end{tablenotes}
    \end{threeparttable}
  \label{tab:prs}
\end{table*}

%Before analyzing the numbers related to the acceptance of the pull requests, it is interesting to check the level of participation of the countries in the projects in general. Figure~\ref{fig:overall} corresponds to the number of pull requests sent by each country among those identified in the 15 projects we analyzed. At first glance we can see that the countries of South America and Africa are mostly in the first scales of the pull requests submissions.

\vspace{0.2cm}
\noindent
\textbf{Acceptance based on the location.}
On average, 32\% of the PRs from developers of all countries were accepted. This number rises to 41\% when we consider just the top-20 countries in Table \ref{tab:prs}. Japan-based developers are the ones with the highest acceptance ratio (51\%) followed by United Kingdom based developers (48\%), and Australia-based developers (46\%). When we look at the how the PR acceptance ratio relates with countries' HDI, we can see that the higher HDI levels tend to have a higher PR acceptance ratio on average, as one can see in Table \ref{tab:hdi}. An exception are the countries grouped at the Low HDI level, which have a PR acceptance ratio more similar to the countries grouped under the Very High HDI level, on average. However, this discrepancy may occur due to the difference between the number of contributors in countries at the Low HDI level (58) and in countries at the Very High HDI level (11,344). This is also observable looking at the number of PRs, which is 110 summing up all Low HDI level countries, while the same number is 36,972 for the Very High HDI level. As a consequence there are countries like Syria, Rwanda, and Senegal with an acceptance rate of 50\%, but with only two PRs. Developers in Africa, for instance, had 52\% of their PRs accepted (although they have performed only 389 pull requests). South Africa based developers, in particular, contributed with 117 of these PRs (with 59\% acceptance rate), although the country has a High HDI (0.705). South American developers faced an even smaller ratio (only 34\% of their 682 contributions were accepted). Moreover, we noted that 8,794 contributors (19\% of the total) performed just one single PR (the so-called drive-by-commits or casual contributors~\cite{pinto2016casuals}, that is, contributors that perform at most one contribution and leave the project). We found that casual contributors are more frequently based in the United States and United Kingdom (43\% of developers based in these two countries performed just one pull request). %Developers from India, China, and Poland are the least casuals (only 25\%, 24\%, and 21\% of them, respectively, performed just one single pull request). 
In a manual inspection of these casual contributions, we found that a significant number of them are related to improving the documentation (e.g., pull request \texttt{18353}\footnote{https://github.com/apple/swift/pull/18353} on \textsc{apple/swift}), although more complex contributions exist, such as the one from a Poland-based contributor who fixed a bug that occurred during the installation of the \textsc{atom/atom} project on Ubuntu linux\footnote{https://github.com/atom/atom/pull/3773}. 

\begin{table}[]
  \caption{HDI vs.~PR acceptance ratio}
  \footnotesize
  \centering
  \begin{tabular}{lrrr}
    \hline
    \multirow{2}{*}{Human Development Index (HDI)}    & \multicolumn{3}{c}{Acceptance Ratio}  \\ 
                            & Median    & Mean      & Std. Dev.     \\ \hline
    Very High (0.800--1.000) & 39.18\%   & 35.46\%   & 15.79\%       \\ 
    High (0.700--0.799)      & 28.57\%   & 29,02\%   & 24,24\%       \\
    Medium (0.550--0.699)    & 20.29\%   & 30,17\%	& 33,63\%       \\
    Low (<0.549)            & 36,36\%   & 32,62\%	& 26,48\%       \\ \hline
  \end{tabular}
  \label{tab:hdi}
\end{table}

%\merge{1 - https://github.com/atom/atom/pull/3773 | Poland | o submitter parece ter indicado uma dependencia de instalação \\ \\
%2 - https://github.com/python/cpython/pull/11162 | China | O submitter parece ter sugerido uma mensagem/alerta para uma exceção (ATUALMENTE ESSE USUÁRIO JA FEZ OUTRAS 2 CONTRIBUIÇÕES PORÉM ELAS FORAM REALIZADAS DEPOIS QUE FIZEMOS A COLETA DE DADOS) \\ \\
%3 - https://github.com/scikit-learn/scikit-learn/pull/12301 | India | Submitter atualizou um link da documentação. Analisando o perfil dele, parece não ter um nível de participação tão grande no github, provavelmente na época do PR poderia estar tentando iniciar suas contribuições em projetos. \\ \\
%4 - https://github.com/apple/swift/pull/18353 | China |  O submitter corrigiu um erro de digitação. o usuário se mostra ativo no github desde 2010. \\ \\
%5 - https://github.com/Microsoft/vscode/pull/57374 | Poland | Submiter abriu uma issue sobre um problema de criação de branch e depois realizou esse pull request pois após resolver o problema de sua issue. Ele não parece ser um usuário muito ativo no github.}

\vspace{0.2cm}
\noindent
\textbf{Rejection based on the location.}
In terms of rejection, it seems that, regardless of their location, having a pull request rejected is commonplace. In particular, 59\% of the overall pull requests were rejected. Interestingly, developers from 29 locations had 100\% of rejections. These contributors, however, made very few contributions (i.e., developers from locations such as Paraguay, Ethiopia, and Burma performed at most seven pull requests). 
When manually inspecting these 100\% rejected pull requests, we noted that some contributors may not yet master how to use Git/GitHub. For instance, pull requests \texttt{12336}\footnote{https://github.com/scikit-learn/scikit-learn/pull/12336} on \textsc{scikit-learn/scikit-learn} does not change a single line of code, and has a misleading commit message.
Moreover, developers from other low HDI locations have contributed more frequently, but still face a high rejection rate. For example, developers based in Indonesia submitted 107 pull requests, with 78 of them rejected (72\%). Bangladesh-based developers submitted 65 pull requests with 87\% of them rejected.
When taking into account only the developers' locations with more than 250 pull requests, we found that Poland-based developers were the ones that faced the most rejections (74\% of their pull requests were rejected), followed by Spain-based developers (72\%), and Brazil-based developers (67\%).
\begin{figure*}[h]
  \includegraphics[width=\linewidth]{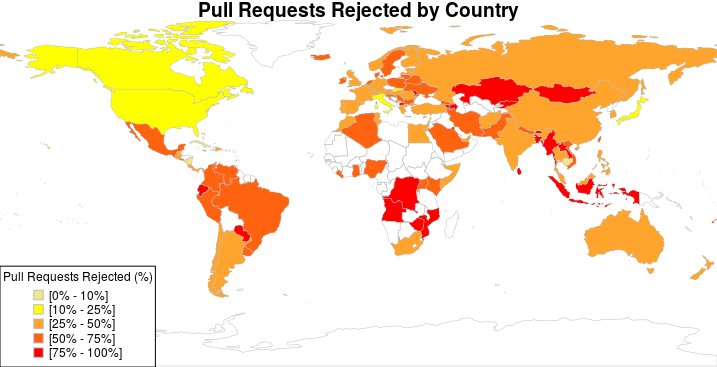}
  \caption{Pull request rejected per developers' location}
  \label{fig:boat1}
\end{figure*}

\section{Related Work}
von Engelhardt~\cite{von2013geographic} employed some heuristics on SourceForge (e.g., email headers, time-zone, and IP address) to infer the location of contributors. 
Bird and colleagues~\cite{Bird:2012:MSR} employed heuristics such as the email domain, social networks, and even commit history to determine the location of the top contributors of Firefox and Eclipse.
Spinellis~\cite{Spinellis:2006:IWGSD} analyzed the FreeBSD operating system by investigating the impact of geographical location on code quality. 
Vasilescu and colleagues~\cite{VasilescuPRBSDF15} used the GitHub location to infer the presence of female developers on OSS projects. Bj\o{}rn and Boulus-R\o{}dje~\cite{Bjorn:2018:CHI} studied the role that infrastructural accessibility plays on the success of tech startup in Palestine. 

To the best of our knowledge, the work of Rastogi et al~\cite{Rastogi:ESEM:2018} is the closest to our work. However, their work focuses on the developers' location that contributed the most. In our study, however, we shed additional light on developers from low HDI locations, which happen to contribute the least or were rejected the most. 

\section{Implications}

Our findings indicate that contributors from lower HDI countries might face a hard time to contribute to open source. Given this observation, open source communities might want to promote sprints, hackathons, and other onboarding programs in these locations. Similarly, companies that fund open source communities might also want to fund mentors in lower HDI locations. These local activities might contribute to foster an open source culture in other less wealthy locations.

\section{Limitations}

First, our study is restricted to GitHub; although GitHub is the largest software development platform, we acknowledge that particular countries might have preferences for other platforms. Similarly, developers in some countries may have low participation in certain popular projects because they do not align with the goals of that country or software developers in that country.

Our study is also limited to the number of pull requests studied, which clearly does not represent all possible forms of contributions available in OSS projects.
Another limitation is that the location field on GitHub is a free form (i.e., it accepts any information). 
Although we employed some additional steps to make sure that the location exists, we still may have considered developers with inaccurate locations (e.g., outdated ones).
Finally, there are many other factors that may influence the pull request decision making process. Our work focused on one factor, the location. Therefore, it is unclear how other factors correlate to ours, which we left for future work.

\section{Conclusions}

In this paper we studied whether the developers' location has any correlation to the pull request decision making. We mined data from 44k pull requests performed in 20 popular OSS projects. We report three main findings: 
First, developers based in high HDI locations, such as United States, United Kingdom, and Germany, are the ones that contribute the most. 
Second, in terms of acceptance, again, developers based in high HDI locations such as Japan and United Kingdom have the highest acceptance ratios.
Third, in terms of rejection, however, developers based in low HDI locations such as Ethiopia, Burma, and Paraguay never had a single contribution accepted. High rejection rates were also common in other low HDI locations.

\bibliographystyle{unsrt}
\bibliography{references}

\begin{thebibliography}{1}

\bibitem{Tsay:2014:IST:2568225.2568315}
Jason Tsay, Laura Dabbish, and James Herbsleb.
\newblock Influence of social and technical factors for evaluating contribution
  in github.
\newblock In {\em ICSE}, pages 356--366, 2014.

\bibitem{un_hdi_report_2019}
United Nations.
\newblock Human development report 2019: Beyond income, beyond averages, beyond
  today: Inequalities in human development in the 21st century.

\bibitem{pinto2016casuals}
Gustavo Pinto, Igor Steinmacher, and Marco~Aur{\'{e}}lio Gerosa.
\newblock More common than you think: An in-depth study of casual contributors.
\newblock In {\em SANER}, pages 112--123, 2016.

\bibitem{von2013geographic}
Sebastian von Engelhardt, Andreas Freytag, and Christoph Schulz.
\newblock On the geographic allocation of open source software activities.
\newblock {\em International Journal of Innovation in the Digital Economy
  (IJIDE)}, 4(2):25--39, 2013.

\bibitem{Bird:2012:MSR}
Christian Bird and Nachiappan Nagappan.
\newblock Who? where? what?: Examining distributed development in two large
  open source projects.
\newblock In {\em MSR}, pages 237--246, 2012.

\bibitem{Spinellis:2006:IWGSD}
Diomidis Spinellis.
\newblock Global software development in the freebsd project.
\newblock In {\em Proceedings of the 2006 International Workshop on Global
  Software Development for the Practitioner}, GSD '06, pages 73--79, 2006.

\bibitem{VasilescuPRBSDF15}
Bogdan Vasilescu, Daryl Posnett, Baishakhi Ray, Mark G.~J. van~den Brand,
  Alexander Serebrenik, Premkumar~T. Devanbu, and Vladimir Filkov.
\newblock Gender and tenure diversity in github teams.
\newblock In {\em CHI}, pages 3789--3798, 2015.

\bibitem{Bjorn:2018:CHI}
Pernille Bj\o{}rn and Nina Boulus-R\o{}dje.
\newblock Infrastructural inaccessibility: Tech entrepreneurs in occupied
  palestine.
\newblock {\em ACM Trans. Comput.-Hum. Interact.}, 25(5), October 2018.

\bibitem{Rastogi:ESEM:2018}
Ayushi Rastogi, Nachiappan Nagappan, Georgios Gousios, and Andr{\'{e}} van~der
  Hoek.
\newblock Relationship between geographical location and evaluation of
  developer contributions in github.
\newblock In {\em ESEM}, pages 22:1--22:8, 2018.

\end{thebibliography}

\begin{IEEEbiographynophoto}{Leonardo B. Furtado}
Is an undergraduate student at the Federal University of Par\'{a}, Brazil. He does research on empirical software engineering.
\end{IEEEbiographynophoto}

\begin{IEEEbiographynophoto}{Bruno Cartaxo}
is an associate professor at the Federal Institute for Education, Science, and Technology of Pernamuco (IFPE), Brazil. 
He received his Ph.D. and a M.Sc. degree in Computer Science from the Center of Informatics (CIn) at Federal University of Pernambuco (UFPE). He conducts pure and applied research in the broad area of Software Engineering and Technology Transfer.
\end{IEEEbiographynophoto}

\begin{IEEEbiographynophoto}{Christoph Treude}
is an ARC DECRA Fellow and a Senior Lecturer in the School of Computer Science at the University of Adelaide, Australia. He received his Ph.D. in computer science from the University of Victoria, Canada in 2012. The goal of his research is to advance collaborative software engineering through empirical studies and the innovation of tools and processes that explicitly take the wide variety of artefacts available in a software repository into account. He currently serves on the editorial board of the Empirical Software Engineering journal and was general co-chair for the IEEE International Conference on Software Maintenance and Evolution 2020.
\end{IEEEbiographynophoto}

\begin{IEEEbiographynophoto}{Gustavo Pinto}
is an assistant professor of computer science at the Federal University of Par\'{a}, Brazil. He received his PhD from Federal University of Pernambuco, Brazil in 2015. His research focuses on the interactions between people and code, spanning the areas of software engineering and programming languages. He currently serves as the co-Editor-in-Chief of the Journal of Software Engineering Research and Development (JSERD). Contact him at gpinto@ufpa.br.
\end{IEEEbiographynophoto}

\end{document}